\documentclass{article}
\usepackage{amssymb}
\usepackage{amsmath}
\usepackage{amsthm}
\usepackage{mathtools}
\usepackage{todonotes}
\usepackage{xurl}
\usepackage[ruled]{algorithm2e}
\usepackage{hyperref}

\usepackage{xcolor}

\newcommand{\Data}{\textbf{D}}
\newcommand{\FunctionFamily}{\textbf{F}}
\newcommand{\Coeffs}{\alpha}
\newcommand{\Labels}{\textbf{Y}}
\newcommand{\Evals}{\textbf{Z}}
\newcommand{\HyperParas}{\textbf{H}}

\newcommand{\Parameters}{\Theta}
\newcommand{\Intermediate}{\tilde{g}}

\title{On Neural Quantum Support Vector Machines}
\author{Lars Simon \\ Bundesdruckerei GmbH \\ {\tt lars.simon@bdr.de} \and Manuel Radons \\ Bundesdruckerei GmbH \\ {\tt manuel.radons@bdr.de}}

\begin{document}
	
	\maketitle

\begin{abstract}
        In \cite{simon2023algorithms} we introduced four algorithms for the training of neural support vector machines (NSVMs) and demonstrated their feasibility. In this note we introduce neural quantum support vector machines, that is, NSVMs with a quantum kernel, and extend our results to this setting. 
\end{abstract}

	\section{Introduction}
	
  The combination of neural networks (NNs) and support vector machines (SVMs) has been explored in numerous publications, see, e.g., \cite{Wiering2013NSVM}, \cite{Qi2016WhenEL}, or \cite{tang2015deep}.
A neural support vector machine (NSVM) is a machine learning model that combines (as the name would suggest) NNs and SVMs in its architecture.
A quantum support vector machine (QSVM) is an SVM with a quantum kernel, see \cite{Schuld_2019hilbert} and \cite{Havl_ek2019supervised}. In this note we introduce neural quantum support vector machines (NQSVMs), which we define to be NSVMs with a quantum kernel.

A key benefit of NSVMs and NQSVMs is that they allow to incorporate and exploit domain knowledge in the model architecture by adapting the NN part to a given data type. 
A straightforward sample application would be to use convolutional neural networks (CNNs) for the design of an N(Q)SVM image classifier.
The key drawback of NSVMs is that the SVM training runtime scales unfavourably with the number of samples in the training set. 

In \cite{simon2023algorithms} we developed four training algorithms for NSVMs that leverage the Pegasos Algorithm to address this problem \cite{Shalev2011Pegasos} and demonstrated their feasibility. 
The main objective of this note is to extend the latter results to NQSVMs.

For fitting QSVMs the Pegasos algorithm is particularly beneficial since there are theoretical bounds on the necessary number of circuit evaluations when using the dual formulation versus using the Pegasos algorithm, cf. \cite{gentinetta2022complexity}.

We note that the first of our four adapted algorithms is very similar to the quantum kernel alignment algorithm via Pegasos in \cite{gentinetta2023quantum}.
Further, 
we would like to stress, as we did in \cite{simon2023algorithms}, that at this point we do not claim any performance improvements over existing training procedures. The scope of this article is restricted to the derivation of the adapted algorithms and a proof of concept. We feel, however, that our experimental results are encouraging and merit further investigation, both theoretical and pratical.\\

\noindent\textbf{Acknowledgement} This article was written as part of the Qu-Gov project, which was commissioned by the German Federal Ministry of Finance. The authors want to extend their gratitude to Manfred Paeschke and Oliver Muth for their continuous encouragement and support.

\subsection{Content and Structure}

In Section \ref{sec:prelim} we cover the necessary preliminaries on SVMs, NSVMs and quantum kernels. Section \ref{sec:algos} contains our main contribution, the description of four NQSVM training algorithms that leverage the Pegasos algorithm.
Our proof of concept experiments are described in Section \ref{sec:experiments}. Section \ref{sec:conclusion} contains our closing remarks.
 
	\section{Preliminaries}
 \label{sec:prelim}
	 In this section we briefly cover some material in preparation of our treatise of neural quantum support vector machines in Section \ref{sec:algos}.
	
	\subsection{Support Vector Machines}
	
	Just like in the work that originally introduced the Pegasos algorithm \cite{Shalev2011Pegasos} we consider support vector machines without bias terms. It is, however, not hard to adapt the algorithms we will present to include a bias term. While it is common to consider the dual formulation of the optimization problem underlying support vector machines, we will only consider the primal formulation; this is appropriate, since the Pegasos algorithm and the algorithms we will present all exclusively make use of the primal formulation. For the sake of brevity we will only state the kernelized version of the primal problem.
	
	Let $n\in\mathbb{Z}_{\geq 1}$, and $m\in\mathbb{Z}_{\geq 2}$. Assume we are given a set of points \[(x_1 , y_1 ),\dots ,(x_m , y_m )\in \mathbb{R}^n\times\{-1,1\}\,.\] This is interpreted as having $m$ datapoints $x_1 ,\dots , x_m\in\mathbb{R}^n$ divided into two classes, where $y_i\in\{-1,1\}$ determines the class of $x_i$, where $i\in\{1,\dots ,m\}$. We further assume that there exists at least one point in each of the two classes, i.e., $\{y_1 ,\dots , y_m\} = \{-1,1\}$. Furthermore, let $K$ be a positive semidefinite kernel on $\mathbb{R}^n$, i.e., a map $K\colon\mathbb{R}^n\times\mathbb{R}^n\to\mathbb{R}$ with $K(a, b) = K(b, a)$ for all $a, b \in\mathbb{R}^n$, such that the matrix \[\left(K(z_i , z_j )\right)_{1\leq i,j\leq l}\] is positive semidefinite, whenever $l\in\mathbb{Z}_{\geq 1}$ and $z_1 , \dots , z_l \in\mathbb{R}^n$. Let $(H, \langle\cdot ,\cdot\rangle_H)$ be the reproducing kernel Hilbert space associated to $K$. The corresponding support vector machine minimization problem then becomes (here, $\lambda\in\mathbb{R}_{>0}$ is a hyperparameter):
	\begin{equation}\label{svm_kernelized_primal}
		\begin{aligned}
			& {\text{minimize}}
			& & \frac{\lambda}{2}{\Vert f\Vert}_H^2 +  \frac{1}{m}\sum_{i=1}^{m}\max \left(0, 1 - y_i f(x_i ) \right)\\
			& \text{subject to}
			& &  f\in H.
		\end{aligned}
	\end{equation}
	
	\subsection{Neural Support Vector Machines}\label{section_nsvm}
	
	Informally speaking, the Neural Support Vector Machine \cite{Wiering2013NSVM} works by first computing features of an input using a neural network and subsequently feeding the so computed features into a support vector machine. The neural network and the support vector machine are trained in unison in order to ensure that the features are maximally relevant.
	
	More formally, let $\left(F_\theta \right)_{\theta\in\mathbb{R}^L}$ be a family of maps $F_\theta\colon\mathbb{R}^d\to\mathbb{R}^n$ and let $K$ be a positive semidefinite kernel on $\mathbb{R}^n$ with associated reproducing kernel Hilbert space $(H, \langle\cdot ,\cdot\rangle_H)$. For simplicity, we further assume that the maps $F\colon\mathbb{R}^d\times\mathbb{R}^L\to\mathbb{R}^n$, $(x, \theta )\mapsto F_\theta (x)$ and $K$ are $\mathcal{C}^\infty$-smooth. In practice, $\left(F_\theta \right)_{\theta\in\mathbb{R}^L}$ is usually chosen to be a neural network, where $\theta$ represents the trainable parameters. The hypothesis class we consider is then the set of functions
	\begin{align*}
		\left\{f\circ F_\theta\colon \mathbb{R}^d\to\mathbb{R}\vert f\in H, \theta\in\mathbb{R}^L\right\}.
	\end{align*}
	Given an element $g$ in this hypothesis class, the associated classifier is given by the map $\mathbb{R}^d\to\{-1,1\}$ which maps $x\in\mathbb{R}^d$ to $1$ whenever $g(x)\geq 0$, and to $-1$ otherwise.
	
	Given data $(x_1 , y_1 ),\dots , (x_m , y_m )\in\mathbb{R}^d\times\{-1,1\}$ with $\{y_1 , \dots , y_m\} = \{-1, 1\}$, we choose a classifier from the hypothesis class by solving the following optimization problem:
	\begin{equation}\label{nsvm_optimization_problem}
		\begin{aligned}
			& {\text{minimize}}
			& & \frac{\lambda}{2}{\Vert f\Vert}_H^2 +  \frac{1}{m}\sum_{i=1}^{m}\max \left(0, 1 - y_i f(F_\theta (x_i )) \right)\\
			& \text{subject to}
			& &  f\in H, \theta\in\mathbb{R}^L,
		\end{aligned}
	\end{equation}
	where, as before, $\lambda >0$ is a hyperparameter.

	\subsection{Quantum Kernels}\label{section_quantum_kernels}
	
	Following \cite{Havl_ek2019supervised}, we consider {\emph{quantum feature maps}} $\Phi\colon\mathbb{R}^n\to\mathcal{M}_N$ of the form
	\begin{align*}
		x\mapsto U(x)|0^N\rangle\langle 0^N|U^\dag (x)
	\end{align*}
	where $n,N\in\mathbb{Z}_{\geq 1}$ and \[U(x)\in\left(\mathbb{C}^{2\times 2}\right)^{\otimes N}\cong \mathbb{C}^{2^N\times 2^N}\] is unitary for all $x\in\mathbb{R}^n$. For simplicity, we assume that \[U\colon\mathbb{R}^n\to\mathbb{C}^{2^N\times 2^N}\] is $\mathcal{C}^\infty$-smooth. Here, $\mathcal{M}_N$ is the $\mathbb{R}$-vector space of Hermitian matrices in $\mathbb{C}^{2^N\times 2^N}$. Of course we have \[\Phi (\mathbb{R}^n )\subseteq\{M\in\mathcal{M}_N | M\text{ positive semidefinite and }\operatorname{Trace}(M)=1\}\,,\] but since the latter set is not an $\mathbb{R}$-vector space with the canonical operations, we consider $\mathcal{M}_N$ instead. Note that $\dim_\mathbb{R} (\mathcal{M}_N) = 4^N$ and that $\mathcal{M}_N$ is an $\mathbb{R}$-Hilbert space when equipped with the Frobenius inner product $\langle\cdot , \cdot\rangle_{\mathcal{F}}$.
	
	It is then clear that the map $K\colon\mathbb{R}^n\times\mathbb{R}^n\to\mathbb{R}$ given by
	\begin{align*}
		K(x,z) & = \langle \Phi(x), \Phi(z)\rangle_{\mathcal{F}}\\
		& = \operatorname{Trace} (\Phi (x)^\dag \Phi(z))\\
		& = \left| \langle 0^N|U^\dag (x) U(z)|0^N\rangle \right|^2
	\end{align*}
	is a $\mathcal{C}^\infty$-smooth positive semidefinite kernel on $\mathbb{R}^n$. Kernels of this form are called {\emph{quantum kernels}}. For evaluation resp.\ estimation of quantum kernels on a quantum computer we refer to \cite{Havl_ek2019supervised} and \cite{glick2022covariant}.

	\subsection{The Reproducing Kernel Hilbert Space of a Quantum Kernel}\label{section_rkhs_of_quantum_kernels}
	Let $K$ be a quantum kernel (we adopt the notation from above). Then the (uniquely determined) reproducing kernel Hilbert space of $K$ is
	\begin{align*}
		H & = \{\operatorname{Trace} (M^\dag \Phi(\cdot ))\colon\mathbb{R}^n\to\mathbb{R} | M\in\mathcal{M}_N\}\\
		& = \left\{
		f\colon\mathbb{R}^n\to\mathbb{R}
		| \exists M\in\mathcal{M}_N\colon f(x) = \langle 0^N|U^\dag (x) M U(x)|0^N\rangle
		\text{ for all }x\in\mathbb{R}^n\right\}
	\end{align*}
	and the norm $\Vert\cdot\Vert_H$ is given as
	\begin{align*}
		\Vert f\Vert_H =
		\inf \{
		\Vert M\Vert_{\mathcal{F}}
		| M\in\mathcal{M}_N\text{ and }f = \operatorname{Trace} (M^\dag \Phi(\cdot )) 
		\}
		\text{ for all }f\in H,
	\end{align*}
	where $\Vert \cdot\Vert_{\mathcal{F}}$ denotes the Frobenius norm. Note that the inner product $\langle\cdot ,\cdot\rangle_H$ on $H$ is uniquely determined by the norm $\Vert\cdot\Vert_H$ via the polarization identities. Furthermore, the canonical mapping $\Gamma\colon\mathcal{M}_N\to H$ given by $M\mapsto \operatorname{Trace} (M^\dag \Phi(\cdot ))$ is $\mathbb{R}$-linear and surjective and, for all $M\in\mathcal{M}_N$, we have $\Vert\Gamma (M)\Vert_H\leq \Vert M\Vert_{\mathcal{F}}$ with equality if and only if $M\in\ker (\Gamma)^\perp$.

	Note that $H$ is also described by the set
	\begin{align*}
		\left\{
		\sum_{i=1}^{l} \alpha_i \left| \langle 0^N|U^\dag (z_i ) U(\cdot )|0^N\rangle \right|^2
		\colon\mathbb{R}^n\to\mathbb{R}
		\bigg| l\in\mathbb{Z}_{\geq 0}, \alpha_i \in\mathbb{R},z_i\in\mathbb{R}^n\,\forall i\in[l]\right\}\,,	\end{align*}
  where $[l]\coloneqq\{1,\dots,l\}$,
	since the latter set is a dense $\mathbb{R}$-vector subspace of $H$ and $\dim_\mathbb{R}H\leq \dim_\mathbb{R}\mathcal{M}_N =4^N<+\infty$.
	
	All of the above is obvious from \cite[Thm. 4.21]{steinwart2008svm}. For a treatment of these facts that is specific to quantum kernels, see also \cite{schuld2021supervised}.

	\section{Neural Quantum Support Vector Machines}\label{sec:algos}
	
	A {\emph{neural quantum support vector machine}} is a neural support vector machine whose kernel is a quantum kernel. So, let $K$ be a quantum kernel and adopt the notation from Sections \ref{section_nsvm}, \ref{section_quantum_kernels}, and \ref{section_rkhs_of_quantum_kernels}.
 
	Roughly speaking, we now replace $H$ by $\mathcal{M}_N$ in optimization problem \ref{nsvm_optimization_problem}. To see why this is appropriate, use the decomposition $\mathcal{M}_N = \ker (\Gamma ) \oplus \ker (\Gamma)^\perp $ and the properties of $\Gamma$ from Section \ref{section_rkhs_of_quantum_kernels}. The optimization problem underlying the neural quantum support vector machine then becomes:
	\begin{equation}\label{nqsvm_optimization_problem}
		\begin{aligned}
			& {\text{minimize}}
			& & \frac{\lambda}{2}{\Vert M\Vert}_{\mathcal{F}}^2 +  \frac{1}{m}\sum_{i=1}^{m}\max \left(0, 1 - y_i \langle 0^N|U^\dag (F_\theta (x_i )) M U(F_\theta (x_i ))|0^N\rangle \right)\\
			& \text{subject to}
			& &  M\in \mathcal{M}_N , \theta\in\mathbb{R}^L\,.
		\end{aligned}
	\end{equation}

\begin{figure}
	\centering
	\begin{tikzpicture}
		\draw [->] (0,-1) -- (0,3); 
		\node[draw] at (2, 2.5)    [align=left] {data\\ space $\mathbb R^d$};
		\draw [->] (-1,0) -- (3,0);
		\draw [thick, ->] (4,1.5) -- (6,1.5) node [midway, above] {NN $F_\theta$};
		\draw [->] (7.5,-1) -- (7.5,3); 
		\draw [->] (6.5,0) -- (10.5,0);
		\node[draw] at (9.5, 2.5)   [align=left] {NN feature\\ space $\mathbb R^n$};
		\node at (1.1,0.9) {\textbullet\ $x$};
		\node at (8.8,0.5) {\textbullet\ $F_\theta(x)$};
		\node at (9.5,-6) {\textbullet\ $\Phi(F_\theta(x))$};
		\node at (-.5,-6) {\textbullet};
		\node[] at (-.5,-5.6) [align=left] {$\operatorname{Tr}(M \Phi (F_\theta(x)))$};
		\node at (1,-6) {\textbullet};
		\node[] at (1.15,-6.2) [align=left] {$0$};
		
		\draw[->,thick] (9,-1.5) -- (9,-3.5) node [midway,left,align=left] {\ \ quantum \\ feature map $\Phi$};
		
		\draw [dotted] (1,-7.5) -- (1,-4.5) node [at start,right,align=left] {decision\\ boundary}; 
		\node[draw] at (2, -4.5) [align=left] {real\\ line $\mathbb R$};
		\draw [->] (-1,-6) -- (3,-6);
		\draw [thick, <-] (4,-5.5) -- (6,-5.5) node [midway, above] {$\operatorname{Tr}(M\cdot (\cdot))$};
		\draw [->] (7.5,-8) -- (7.5,-4); 
		\draw [->] (6.5,-7) -- (10.5,-7);
		\node[draw] at (9.5, -4.5)   [align=left] {QSVM feature\\ space $\mathcal M_N$};
		
	\end{tikzpicture}
	\caption{This figure indicates how a trained neural quantum support vector machine makes predictions.}
	\label{fig:nqsvm}
\end{figure}
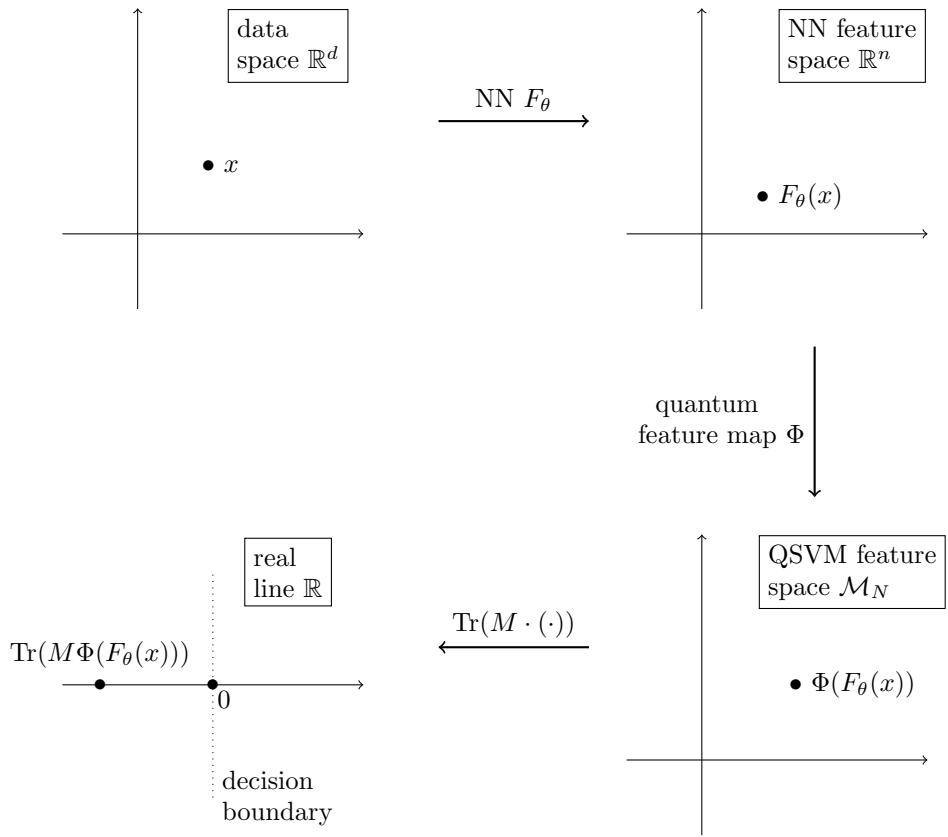

For a visual representation, see Figure \ref{fig:nqsvm}. We are now ready to state the four announced quantum-analogues of the NSVM training algorithms described in \cite{simon2023algorithms}.
Since the use of quantum kernels does not affect the validity of the algorithms' derivation, we will merely state an updated pseudocode of their quantum versions and refer to the latter reference for technical justifications.  

All algorithms take as an input the number of training steps $T\in\mathbb{Z}_{\geq 1}$, a family of maps $\FunctionFamily\coloneqq \left(F_\theta \right)_{\theta\in\mathbb{R}^L}$, where $F_\theta\colon\mathbb{R}^d\to\mathbb{R}^n$, a dataset \[\Data\coloneqq\{(x_1 , y_1 ),\dots , (x_m , y_m )\}\subset\mathbb{R}^d\times\{-1,1\}\] with $\{y_1 , \dots , y_m\} = \{-1, 1\}$, and
a $\mathcal{C}^\infty$-smooth map $U\colon\mathbb{R}^n\to\mathbb{C}^{2^N\times 2^N}$, where $n,N\in\mathbb{Z}_{\geq 1}$, with $U(z)$ unitary for all $z\in\mathbb{R}^n$, which, in practice, is usually
 given as a parametrized $N$-qubit quantum circuit. 
 Further, the inputs will include hyperparameters that vary between the algorithms. 
 For the sake of simplicity we assume that the map $F\colon\mathbb{R}^d\times\mathbb{R}^L\to\mathbb{R}^n$, $(x, \theta )\mapsto F_\theta (x)$ is $\mathcal{C}^\infty$-smooth.

From the output of each algorithm we construct a function $g: \mathbb R^d\to\mathbb R$ that induces a classifier 
\[
  \mathbb{R}^d\to\{-1, 1\}\,,\quad
	x\mapsto
	\begin{cases}
		1 & \text{if }g(x)\geq 0\,,\\
		-1 & \text{if }g(x)<0\,.
	\end{cases}
\]
The construction method for $g$ varies from algorithm to algorithm.

We note that all algorithms will involve gradient descent steps wrt.\ the trainable parameters $\theta$; for implementation details we refer to \cite{simon2023algorithms}.  
Gradients of the quantum kernel are estimated. Since finite difference approximation is computationally expensive, we instead estimate gradients of quantum kernels using the gradient estimator featuring in the well-known SPSA algorithm \cite{spall1992spsa}. While the gradients for the layers of the neural network can be calculated in the usual way, they are still affected by the use of the aforementioned gradient estimator (because of backpropagation).  

Moreover, the vector of coefficients $\alpha$ will, in general, be sparse. For brevity we refrain from mentioning this fact in the pseudocode, even though its exploitation is a necessity in any sensible implementation.

\subsection{Remarks on the Algorithms}

As noted above, the derivation of Algorithms \ref{algo:1}-\ref{algo:4} is layed out in detail in \cite{simon2023algorithms}.
However, some remarks are in order. 

Algorithm 1 is very similar to the algorithm presented in \cite{gentinetta2023quantum}.
The key difference between Algorithms \ref{algo:1} and \ref{algo:2} lies in the update step for $\theta$.
In Algorithm \ref{algo:2}, for step $t\in\{1,\dots T\}$, we first update $f_t$ with a gradient descent step as in Algorithm \ref{algo:1}, to obtain some $\tilde{f}_{t+1}\in H$. Then $\theta_t$ is updated with a gradient descent step to obtain some $\theta_{t+1}\in\mathbb{R}^L$. Subsequently, we obtain $f_{t+1}$ by "approximately projecting" $\tilde{f}_{t+1}$ to $\operatorname{span}_\mathbb{R}(\{ K( F_{\theta_{t+1}} (x_i ) , \cdot ) \vert i\in\{1,\dots ,m\}\})$.
Algorithm \ref{algo:3} introduces mini batching and a more involved minimization objective that is utilized in the gradient descent step. 

Finally, Algorithm \ref{algo:4} consists of two parts: First we train the neural network with the aim of improving the kernel alignment, then we fit a support vector machine. Since the parameters of the neural network are no longer being trained when we fit the support vector machine, we can choose {\emph{any}} algorithm for fitting a support vector machine, e.g. the Pegasos algorithm, or an algorithm making use of the dual formulation. In order to accomodate this freedom of choice, we will leave the second part of the pseudocode somewhat open, which of course leads to a less strict description of the algorithm.

For the construction of the above-mentioned function $g\colon\mathbb{R}^d\to\mathbb{R}$ from the output of the respective algorithms, we again refer to \cite{simon2023algorithms}.

	\begin{algorithm}
		\SetKwInOut{Input}{Input}
		\SetKwInOut{Output}{Output}
		
		\Input{dataset \Data, number of steps $T$, hyperparameter $\lambda$, parametrized unitary $U$, family of functions $\FunctionFamily$}
		\Output{coefficients $\Coeffs$, labels $\Labels$, evaluations $\Evals$, parameters 
  $\Parameters$}
		
		Initialization: randomly pick $\theta_1\in\mathbb{R}^L$ and $i_1\in\{1,\dots , m\}$ and set up a parametrized quantum circuit for evaluating 
  \[K\colon\mathbb{R}^n\times\mathbb{R}^n\to\mathbb{R}\,,\quad K(a,b) = \left| \langle 0^N|U^\dag (a) U(b)|0^N\rangle \right|^2\]
  
  $\alpha_1 = 1$

  $z_1 = F_{\theta_1}(x_{i_1})$

  $\theta_2 = \theta_1$

        \For{t=2,\dots, T}
        {
Choose $i_t\in\{1,\dots , m\}$ uniformly at random
		
  $z_t = F_{\theta_t}
(x_{i_t})\in\mathbb{R}^n$

Compute \[\Intermediate_t = \frac{1}{\lambda (t-1)}\sum_{s=1}^{t-1}\alpha_s y_{i_s} K(z_s , z_t )\]

 \If{$y_{i_t}\Intermediate_t < 1$}{
 $\alpha_t = 1$

 Obtain $\theta_{t+1}\in\mathbb{R}^L$ from $\theta_t$ via a gradient descent step wrt.\ the minimization objective \[
 \mathbb{R}^L\to\mathbb{R},\quad \theta\mapsto -\frac{y_{i_t}}{\lambda (t-1)}\sum_{s=1}^{t-1}\alpha_s y_{i_s} K(z_s , F_\theta (x_{i_t}) )\]
 (gradients of $K$ are estimated using the SPSA-estimator and backpropagated the usual way)
 }
        \Else{
        $\alpha_t  =0$
        
		$\theta_{t+1} = \theta_t$
        }

        }
		
  Set $\alpha\coloneqq(\alpha_1 , \dots , \alpha_T )$

  Set $\Labels\coloneqq\left(y_{i_1},\dots , y_{i_T}\right)$

  Set $\Evals\coloneqq (z_1 , \dots , z_T )$

  Set $\Parameters\coloneqq\theta_{T+1}$

		\Return ($\alpha, \Labels, \Evals, \Parameters$)
		
		\caption{}
		\label{algo:1}
	\end{algorithm}

\begin{algorithm}
		\SetKwInOut{Input}{Input}
		\SetKwInOut{Output}{Output}
		
		\Input{dataset \Data, number of steps $T$, hyperparameter $\lambda$, parametrized unitary $U$, family of functions $\FunctionFamily$}
		\Output{coefficients $\Coeffs$, evaluations $\Evals$, parameters 
  $\Parameters$}
		
		Initialization: randomly pick $\theta_1\in\mathbb{R}^L$ and $i_1\in\{1,\dots , m\}$ and set up a parametrized quantum circuit for evaluating 
  \[K\colon\mathbb{R}^n\times\mathbb{R}^n\to\mathbb{R}\,,\quad K(a,b) = \left| \langle 0^N|U^\dag (a) U(b)|0^N\rangle \right|^2\]

  Let $\alpha^{(2)}\in\mathbb{R}^m$ with $\alpha_{i_1}^{(2)}=1$ and $\alpha_j^{(2)}=0\ \forall\, j\in\{1,\dots ,m\}$ with $j\neq i_1$

  $\theta_2=\theta_1$

        \For{t=2,\dots, T}
        {
Choose $i_t\in\{1,\dots , m\}$ uniformly at random

For all $j\in\{1,\dots ,m\}$ with $j\neq i_t$: set $\alpha_j^{(t+1)}=\alpha_j^{(t)}$

Compute
		\[\Intermediate_t = \frac{1}{\lambda (t-1)}\sum_{j=1}^{m}\alpha_j^{(t)} y_j K(F_{\theta_t} (x_j ) , F_{\theta_t} (x_{i_t}) )
		\]

 \If{$y_{i_t}\Intermediate_t < 1$}{
 $\alpha_{i_t}^{(t+1)} = \alpha_{i_t}^{(t)} + 1$

 Obtain $\theta_{t+1}\in\mathbb{R}^L$ from $\theta_t$ via a gradient descent step wrt.\ the minimization objective
 \[\mathbb{R}^L\to\mathbb{R}\,,\quad \theta\mapsto -\frac{y_{i_t}}{\lambda (t-1)}\sum_{j=1}^{m}\alpha_j^{(t)} y_j K(F_{\theta} (x_j ) , F_{\theta} (x_{i_t}) )\]
 (gradients of $K$ are estimated using the SPSA-estimator and backpropagated the usual way)
 }
        \Else{
       $\alpha_{i_t}^{(t+1)}   =\alpha_{i_t}^{(t)}$
       
	$\theta_{t+1} = \theta_t$
        }

        }
        \For{j=1,\dots,m}{$z_j \coloneqq F_{\theta_{T+1}}(x_j )$}
  
  Set $\alpha\coloneqq\alpha^{(T+1)}$

  Set $\Evals\coloneqq (z_1 , \dots , z_m )$

  Set $\Parameters\coloneqq\theta_{T+1}$

		\Return ($\alpha, \Evals, \Parameters$)
		
		\caption{}
		\label{algo:2}
	\end{algorithm}

\begin{algorithm}
		\SetKwInOut{Input}{Input}
		\SetKwInOut{Output}{Output}
		
		\Input{dataset \Data, number of steps $T$, hyperparameters $\lambda ,\mu$, parametrized unitary $U$, family of functions $\FunctionFamily$, batch size $k$}
		\Output{coefficients $\Coeffs$, evaluations $\Evals$, parameters 
  $\Parameters$}
		
		Initialization: randomly pick $\theta_1\in\mathbb{R}^L$ and set up a parametrized quantum circuit for evaluating 
  \[K\colon\mathbb{R}^n\times\mathbb{R}^n\to\mathbb{R}\,,\quad K(a,b) = \left| \langle 0^N|U^\dag (a) U(b)|0^N\rangle \right|^2\]
		
		Choose a subset $A_1$ of $\{1,\dots , m\}$ with cardinality $k$ uniformly at random
		
		Let $\alpha^{(2)}\in\mathbb{R}^m$ with $\alpha_{j}^{(2)}=\frac1k$ if $j\in A_1$ and $\alpha_j^{(2)}=0$ otherwise 
		
		$\theta_2=\theta_1$

        \For{t=2,\dots, T}
        {
Choose a subset $A_t$ of $\{1,\dots , m\}$ with cardinality $k$ uniformly at random

For all $j\in\{1,\dots ,m\}$ with $j\not\in A_t$: set $\alpha_j^{(t+1)}=\alpha_j^{(t)}$

For all $i\in A_t$ compute
		\[\Intermediate^{(i)}_t = \frac{1}{\lambda (t-1)}\sum_{j=1}^{m}\alpha_j^{(t)} y_j K(F_{\theta_t} (x_j ) , F_{\theta_t} (x_i ) ) 
		\]

\For{all $i\in A_t$}{
\If{$y_i \Intermediate^{(i)}_t < 1$}{$\alpha_{i}^{(t+1)} = \alpha_{i}^{(t)} + \frac{1}{k}$}
\Else{$\alpha_{i}^{(t+1)} = \alpha_{i}^{(t)}$}
}

Obtain $\theta_{t+1}\in\mathbb{R}^L$ from $\theta_t$ via a gradient descent step wrt.\ the minimization objective $\mathbb{R}^L\to\mathbb{R}$: 
					\begin{align*}
						\theta\mapsto & \mu\cdot\frac{
							\sum_{i,j\in A_t} \alpha^{(t+1)}_i \alpha^{(t+1)}_j y_i y_j K(F_\theta (x_i ), F_\theta (x_j ))
						}
					   {
					   	   \sqrt{\sum_{i,j\in A_t} \left( \alpha^{(t+1)}_i \alpha^{(t+1)}_j\right)^2}
					   	   \cdot
					   	   \sqrt{\sum_{i,j\in A_t} K(F_\theta (x_i ), F_\theta (x_j ))^2}
				   	   }\\
						& -
						\frac{
							\sum_{i,j\in A_t} y_i y_j K(F_\theta (x_i ), F_\theta (x_j ))
						}
						{
							k
							\cdot
							\sqrt{\sum_{i,j\in A_t} K(F_\theta (x_i ), F_\theta (x_j ))^2}
						}
					\end{align*}
                    (gradients of $K$ are estimated using the SPSA-estimator and backpropagated the usual way)
                    
                    If $\alpha^{(t+1)}_i = 0$ for all $i\in A_t$, then the first fraction is not well-defined. In this case we replace the first fraction by $0$

}
        \For{j=1,\dots,m}{$z_j \coloneqq F_{\theta_{T+1}}(x_j )$}
  
  Set $\alpha\coloneqq\alpha^{(T+1)}$

  Set $\Evals\coloneqq (z_1 , \dots , z_m )$

  Set $\Parameters\coloneqq\theta_{T+1}$

		\Return ($\alpha, \Evals, \Parameters$)
		
		\caption{}
		\label{algo:3}
	\end{algorithm}

\begin{algorithm}
		\SetKwInOut{Input}{Input}
		\SetKwInOut{Output}{Output}
		
		\Input{dataset \Data, number of steps $T$, parametrized unitary $U$, family of functions $\FunctionFamily$, batch size $k$, loss function $\mathcal{L}$, hyperparameters $\HyperParas$}
		\Output{classifier $h$, parameters 
  $\Parameters$}
		
		Initialization: Randomly pick $\theta_1\in\mathbb{R}^L$ and set up a parametrized quantum circuit for evaluating 
  \[K\colon\mathbb{R}^n\times\mathbb{R}^n\to\mathbb{R}\,,\quad K(a,b) = \left| \langle 0^N|U^\dag (a) U(b)|0^N\rangle \right|^2\] 

\textbf{Part 1} (kernel alignment):

        \For{t=1,\dots, T}
        {
Choose a subset $A_t$ of $\{1,\dots , m\}$ with cardinality $k$ uniformly at random

Obtain $\theta_{t+1}\in\mathbb{R}^L$ from $\theta_t$ via a gradient descent step wrt.\ the minimization objective $\mathbb{R}^L\to\mathbb{R}$: 
				\begin{align*}
					\theta\mapsto \mathcal{L}\left(1,
					\frac{
						\sum_{i,j\in A_t} y_i y_j K(F_\theta (x_i ), F_\theta (x_j ))
					}
					{
						k
						\cdot
						\sqrt{\sum_{i,j\in A_t} K(F_\theta (x_i ), F_\theta (x_j ))^2}
					}\right)
				\end{align*} (gradients of $K$ are estimated using the SPSA-estimator and backpropagated the usual way)
}
\For{$j=1,\dots,m$}{Set $z_j := F_{\theta_{T+1}}(x_j )$}

\textbf{Part 2} (fit SVM):

Fit SVM with kernel $K$ on data $(z_1,y_1),\dots, (z_m,y_m)\in\mathbb R^n\times\{-1,1\}$, using $\HyperParas$ and obtain classifier  \[h\colon\mathbb{R}^n\to\{-1,1\}\]

  Set $\Parameters\coloneqq\theta_{T+1}$

		\Return ($h, \Parameters$)
		
		\caption{}
		\label{algo:4}
	\end{algorithm}

\subsection{Comparison with Quantum Support Vector Machines with Parametrized Kernel}

The performance of a QSVM is closely connected to the suitability of its kernel for a given dataset. 
For example, in \cite{Liu_2021} the authors constucted a binary classification problem which is closely related to the discrete logarithm problem (DLP), on which the security of cryptographic key exchange protocols such as Diffie-Hellman and ElGamal is based, because it is believed to be computationally hard.
A quantum kernel was specifically tailored to the underlying structure of the data  set (by using parts of Shor's algorithm for DLP \cite{Shor_1997} as a subroutine), thus establishing a quantum advantage for the corresponding QSVM over classical learners under the assumption that DLP is indeed classically intractable. 
On this data set, one should not expect such an advantage for QSVMs with generic quantum kernels.

In general, it is not obvious how to construct quantum kernels that exploit the intrinsic structure of a given data set. Moreover, in practice one often has limited knowledge about the data structure.
In \cite{glick2022covariant}, the above observations are explained in more detail and serve as motivation to introduce an approach to finding quantum kernels which are well suited for a given dataset: The approach is to consider a parametrized quantum kernel $(K_\tau )_{\tau\in \mathbb{R}^p}$ and to choose an appropriate $\tau$ using a classical optimization loop.

Strictly speaking, via the definition $K_\theta (\cdot , \cdot ) := K(F_\theta (\cdot ), F_\theta (\cdot ))$, the neural quantum support vector machine can be seen as a special case of a quantum support vector machine with parametrized quantum kernel. However, in the literature on parametrized quantum kernels that the authors are aware of, the parameters $\tau$ only enter the picture via parametrized gates resp.\ unitaries $V_\tau$ (which are independent from the kernel inputs), i.e., the data is {\emph{explicitly not}} passed through a neural network prior to being fed into the kernel. When adjusting the trainable parameters $\tau$ of such a parametrized quantum kernel during training, one changes the mapping of the data to the quantum feature space $\mathcal{M}_N$, but does not transform the data itself prior to feeding it into the quantum kernel. In contrast, when adjusting the trainable parameters $\theta$ of a neural quantum support vector machine during training, the quantum kernel (and thus the mapping of the data to the quantum feature space $\mathcal{M}_N$) stays unchanged, but we change how the data is transformed prior to being fed into the kernel. So, roughly speaking, one can say that a quantum support vector machine with parametrized kernel (where the parameters only enter the picture via parametrized gates which are independent from the kernel inputs) works by modifying the kernel during training to better suit the data, whereas a neural quantum support vector machine modifies the data transformation during training to better suit the (fixed) kernel.
	
On the other hand, it is certainly possible to consider a parametrized kernel $(K_\tau )_{\tau\in\mathbb{R}^p}$ in the neural (quantum) support vector machine, i.e., to consider $\mathcal{K}_{(\theta ,\tau )} (\cdot , \cdot ) := K_\tau (F_\theta (\cdot ), F_\theta (\cdot ))$. The four algorithms we present can trivially be adapted to this setting by optimizing over the concatenated parameter vector $(\theta ,\tau )\in\mathbb{R}^{L+p}$ instead of over $\theta$.

\section{Experiments}
\label{sec:experiments}
The purpose of our numerical experiments is to demonstrate the feasibility of the above adaptation of the algorithms described in \cite{simon2023algorithms} to the quantum setting.
As in the latter reference, we neither extensively tune hyerparameters, etc., to optimize performance, nor do we exhaustively benchmark the performance of our algorithms when compared to each other or to other well-known machine learning models. 
 
 As before, we showcase the incorporation of domain knowledge in N(Q)SVMs by choosing the neural layers depending on the characteristics of the data by training a CNN in combination with the ZZFeatureMap on the MNIST dataset, cf. \cite{lecun2010mnist}, and on the Fashion-MNIST dataset, cf. \cite{xiao2017fashionmnist}.

 All quantum circuits were simulated using the AerSimulator provided by the Qiskit framework. Each circuit run was executed with $450$ shots; this includes circuit runs that were used to approximate/estimate gradients.
Gradients of the quantum kernel were estimated using the gradient estimator featuring in the SPSA algorithm. While the gradients for the layers of the neural network were computed in the usual way, they were also affected by the estimation of the gradients for the quantum kernel, since the gradient estimates for the quantum kernel were, of course, fed to the neural network during backpropagation.

	\subsection{MNIST Data Set}

As we merely aimed for a proof of concept we restricted our experiments to the binary classification task of distinguishing the digits $0$ and $1$.
Adapting our algorithms to multiclass classification and regression problems is a straighforward task that we leave to future work.

Our training and testing set contain $12665$, resp., $2115$ $28\times 28$ grayscale images. Both are approximately balanced between digits $0$ and $1$.
$500$ images from each class (so, a total of $1000$ images) were split off from the training set to serve as a validation set during hyperparameter tuning. After deciding on a hyperparameter configuration, we trained the models from scratch on the entire training set (i.e., including the $1000$ images that were previously split off).

\subsubsection{Description of Models}

All algorithms use the same model, a convolutional layer with $10\times 10$ filters and $4$ output channels, followed by channel-wise dropout, followed by a $2\times 2$ max pooling layer and a subsequent ReLU activation.
The output is then fed into a fully connected layer with input dimension $324$ and output dimension $4$. The output of the latter is normalized with respect to the Euclidean norm in $\mathbb{R}^{4}$ (where the denominator is artificially bounded from below by a small $\epsilon >0$ in order to avoid division by $0$ and to provide numerical stability) and subsequently scaled by the factor $2$.
Afterwards, the $\tanh$ activation function is applied component-wise and its output is scaled by the factor $\frac{\pi}{4}$. 

As the {\emph{Quantum Feature Map}} we use the above-mentioned ZZFeatureMap with feature dimension $4$, with $1$ repetition, and with full entanglement, cf. \cite{Havl_ek2019supervised}).

\subsubsection{Training and Results}
All four algorithms only needed a relatively small number of training steps; as a result, none of the models encountered all entries of the training set during training. 

All algorithms used gradient descent with momentum and $L^2$-regularization for the $\theta$-update (Algorithm \ref{algo:4} in Part I, the kernel alignment).
The hyperparameter $\lambda$ was set to $0.0001$ in all algorithms (in Algorithm 4 during Part II, the QSVM fit via Pegasos). Algorithm 3 had an additional hyperparameter $\mu = 1$.
The batch size of Algorithms \ref{algo:3} and \ref{algo:4} was $4$.
The loss function during Part I of Algorithm \ref{algo:4} was
$\mathcal{L}(\beta , \gamma ) = (\beta - \gamma )^2$.

Algorithms \ref{algo:1} and \ref{algo:2} achieved 99.1\% and 99.5\% classification accuracy on the test set, respectively, each after 1200 steps.
Algorithm \ref{algo:3} achieved 99.9\% classification accuracy after 600 steps.
Algorithm \ref{algo:4} achieved 99.7\% classification accuracy after 1200 steps, which were composed of 600 steps in Part I and 600 steps in Part II.

	\subsection{Fashion-MNIST Data Set}

Similiarly to before we restrict our experiments to the binary classification task of distinguishing between images of pullovers and sandals.

Our training and testing set contain $12000$, resp., $2000$ $28\times 28$ grayscale images. Both are perfectly balanced between images of pullovers and images of sandals.
$750$ images from each class (so, a total of $1500$ images) were split off from the training set to serve as a validation set during hyperparameter tuning.

\subsubsection{Description of Models}

In order to demonstrate the flexibility of our algorithms we used the same NQSVM models for the experiments with the Fashion-MNIST data set as for the experiments with the MNIST data set.

\subsubsection{Training and Results}
All four algorithms only needed a relatively small number of training steps; as a result, none of the models encountered all entries of the training set during training. 

The training procedure for the experiments with the Fashion-MNIST data set was the same as the training procedure for the experiments with the MNIST data set, except that we chose different values for learning rate and momentum for the $\theta$-update, as well as different numbers of steps for the respective algorithms.

Algorithms \ref{algo:1} and \ref{algo:2} achieved 98.7\% and 98.0\% classification accuracy on the test set, respectively, each after 1200 steps.
Algorithm \ref{algo:3} achieved 98.7\% classification accuracy after 2000 steps.
Algorithm \ref{algo:4} achieved 99.3\% classification accuracy after 3000 steps, which were composed of 1500 steps in Part I and 1500 steps in Part II.

We would like to stress that we do not claim superior performance in comparison to the state of the art. 
After all, both classifications problems we considered in our experiments were easy. The point of this section was merely to provide a proof of concept that we can successfully train a NQSVM with convolutional layers and quantum kernel using Algorithms \ref{algo:1}-\ref{algo:4}.

	\section{Conclusion}\label{sec:conclusion}
The goal of this article was to incorporate quantum kernels in the N(Q)SVM training algorithms derived in \cite{simon2023algorithms} and to provide a proof of concept for this approach \emph{without} any claims of superiority over the current state of the art. 
However, as in the case of the latter work, which had a similarly modest goal, we think that the numerical results are encouraging and merit further investigation.
A list of candidates for future work includes, but is not restricted to, the move from binary classification to broader learning tasks, a thorough theoretical analysis of the algorithms (with respect to convergence properties and computational costs, etc.) and the research of relevant use cases where our training methodology displays a tangible advantage over existing approaches.

\section*{Declarations}
\subsection*{Competing interests}
There are no competing interests. Both authors work for the 
same institution.
\subsection*{Authors' contributions} 
The first author developed the theory, the second author wrote the paper. Experiments were devised in cooperation. The first author implemented them. 
\subsection*{Funding}
No funding was received. 
\subsection*{Availability of data and materials} 
The MNIST data set \cite{lecun2010mnist} is available at 
\url{http://yann.lecun.com/exdb/mnist}\,. The Fashion-MNIST data set \cite{xiao2017fashionmnist} is available at \url{https://github.com/zalandoresearch/fashion-mnist}\,.

\bibliographystyle{alpha}
\bibliography{references-all}
\end{document}